\documentclass[10pt]{article}
\usepackage{amsmath}
\usepackage{graphicx}
\usepackage{amsfonts}
\usepackage{amssymb}
\usepackage{epsf}
\usepackage{latexsym}

\def\80{\hspace{0.8in}}

\def\bq{\bf q}
\def\bR{\bf R}
\def\br{\bf r}
\def\brho{\mbox{\boldmath$\rho$}}
\newcommand{\be}{\begin{enumerate}}
\newcommand{\ee}{\end{enumerate}}
\newcommand{\bi}{\begin{itemize}}
\newcommand{\ei}{\end{itemize}}
\newcommand{\bd}{\begin{description}}
\newcommand{\ed}{\end{description}}

\def\beq{\begin{equation}}
\def\eeq{\end{equation}}
\def\bea{\begin{eqnarray}}
\def\eea{\end{eqnarray}}
\def\d{\textrm{d}}

\def\cr{\mbox{\scriptsize{\bf $\mbox{ } \times \mbox{ }$}}}

\def\mj{\mbox{j}}

\def\mn{\mbox{n}}

\def\mz{\mbox{z}}

\def\mC{\mbox{C}}

\def\mZ{\mbox{Z}} 
\def\sa{\mbox{\scriptsize a}}
\def\sb{\mbox{\scriptsize b}}
\def\sc{\mbox{\scriptsize c}}

\def\sC{\mbox{\scriptsize C}}

\def\sF{\mbox{\scriptsize F}}

\def\sH{\mbox{\scriptsize H}}

\def\sK{\mbox{\scriptsize K}}

\def\sN{\mbox{\scriptsize N}}

\def\sS{\mbox{\scriptsize S}}
\def\sT{\mbox{\scriptsize T}}

\def\eph(B){\mbox{\scriptsize emergent(LMB)}}

\def\tH{\mbox{\tiny H}}

\def\fE{\mbox{\sffamily E}}

\def\fR{\mbox{\sffamily R}}
\def\fS{\mbox{\sffamily S}}
\def\fT{\mbox{\sffamily T}}
\def\fU{\mbox{\sffamily U}}

\def\bn{\mbox{\bf n}}

\begin{document}
\begin{titlepage}
\vspace{.7in}
\begin{center}
 
\vspace{2in} 

{\LARGE\bf SHAPE QUANTITIES FOR}

\vspace{.1in}

{\LARGE\bf RELATIONAL QUADRILATERALLAND}

\vspace{.2in}

{\bf Edward Anderson}$^{1}$ 

\vspace{.2in}

{\em $^1$ DAMTP Cambridge U.K.} and 

{\em Departamento de F\'{\i}sica Te\'{o}rica, Universidad Autonoma de Madrid.}  

\vspace{.2in}

\end{center}

\begin{abstract}

I investigate the question of what is the relational quadrilateral counterpart of the Dragt-type coordinates that are useful for the relational triangle. 
These relational particle models are relational in the sense that only relative times, relative ratios of separations and relative 
angles are significant.
These models have many analogies with the geometrodynamical formulation of general relativity.  
This renders their classical and quantum mechanics study as suitable toy models for 
1) studying Problem of Time in Quantum Gravity strategies, 
in particular timeless, semiclassical and histories theory approaches and combinations of these. 
2) For consideration of various other quantum-cosmological issues, such as structure 
formation/inhomogeneity and notions of uniform states and their significance.  
The relational quadrilateral is more useful in these respects than previously investigated simpler 
RPM's due to simultaneously possessing linear constraints, nontrivial subsystems and the nontrivial 
complex-projective mathematics that is characteristic of the general relational $N$-a-gon.

\end{abstract}

\vspace{0.2in}

Seminar I (of three) on relational quadrilaterals.  


\vspace{0.2in}

\end{titlepage}

\section{Introduction}

Relational particle mechanics (RPM) are mechanics in which only relative times, relative angles and 
(ratios of) relative separations are physically meaningful.  
Scaled RPM and pure-shape (i.e. scalefree, and so involving just ratios) RPM were set up 
respectively in \cite{BB82} by Barbour--Bertotti and in \cite{B03} by Barbour.  
These theories are relational in Barbour's sense rather than Rovelli's \cite{Rovellibook, B94I, EOT, 
08I}, via obeying the following postulates. 
Firstly, RPM's are {\it temporally relational} \cite{BB82, RWR, FORD}, i.e. there is no meaningful 
primary notion of time for the whole (model) universe. 
This is implemented by using actions that are manifestly reparametrization invariant while also being 
free of extraneous time-related variables [such as Newtonian time or the lapse in General Relativity 
(GR)].   
This reparametrization invariance then directly produces primary constraints quadratic in the momenta 
(which in the GR case is the super-Hamiltonian constraint, and for RPM's is an energy constraint.) 
Secondly, RPM's are {\it configurationally relational}, which can be thought of in terms of a certain 
group $G$ of transformations that act on the theory's configuration space $Q$ being physically 
meaningless \cite{BB82, RWR, FORD, FEPI, Cones}.   
Configurational relationalism can be approached by \cite{BB82, B03} indirect means that amount to 
working on the principal bundle P($Q$, $G$).  
Then there are linear constraints from varying with respect to the $G$-generators (the super-momentum 
constraint in the case of GR, zero total angular momentum for the universe for scaled RPM and both 
that, and its dot- rather than cross-product counterpart, zero total dilational momentum, for pure-shape 
RPM).   
Then elimination of the $G$-generator variables from the Lagrangian form of these constraints sends one 
to the desired quotient space\footnote{Here, 
$Q$ is the na\"{\i}ve configuration spaces for $N$ particles in dimension $d$, $\mathbb{R}^{Nd}$. 
$G$ is a group of transformations that are declared to be physically meaningless.  
E.g. 1) For scaled RPM, it is the Euclidean group of translations and rotations (removing what is 
usually regarded as the absolutist content of mechanics). 
E.g. 2) For pure-shape RPM, it is the similarity group that furthermore includes dilations.}  
$Q/G$.
In 1- and 2-$d$, these RPM's can be obtained alternatively by working directly on $Q/G$ 
\cite{FORD, Cones}. 
This involves applying a `Jacobi--Synge' construction of the natural mechanics associated with a 
geometry, to Kendall's shape spaces \cite{Kendall} for pure-shape RPM's or to the cones over these 
\cite{Cones} for the scaled RPM's.

RPM's are furthermore motivated through being useful toy models \cite{K92, EOT, RWR, 
Kieferbook06I06IISemiclIGrybBanalMGM, Records, FEPI, AF, Cones} of GR in geometrodynamical form, since 
they possess analogues of many features of this.
RPM's are similar in complexity and in number of parallels with GR to minisuperspace \cite{Mini} or   
2 + 1 GR \cite{Carlip} are, though each such model differs in how it resembles GR, so that such models 
offer complementary perspectives and insights.
In particular, RPM's purely quadratic energy constraint parallels the GR super-Hamiltonian constraint 
in leading to the frozen formalism facet of the Problem of Time \cite{K92, I93}, alongside nontrivial 
linear constraints that parallel the GR momentum constraint (and are absent for minisuperspace), which 
cause a number of further difficulties in various approaches to the Problem of Time.    
Other useful applications of RPM's not covered by minisuperspace models include that RPM's are useful 
for the qualitative study of the quantum-cosmological origin of structure formation/inhomogeneity 
(scaled RPM's are a tightly analogous, simpler version of Halliwell--Hawking's \cite{HallHaw} model 
for this), and for the study of correlations between localized subsystems of a given instant. 
Scalefree RPM's such as this paper's, moreover, occur as subproblems within scaled RPM's, 
corresponding to the light fast modes/inhomogeneities of shape in the preceding sentence's context, 
and so are a useful part of the study of the scaled model too.  
RPM's also admit analogies with conformal/York-type \cite{York73York74ABFOABFKO} initial value 
problem formulations of GR, with 
the conformal 3-geometries playing an analogous role to the pure shapes \cite{York73York74ABFOABFKO}.  
In general, having multiple toy models for the Problem of Time/Quantum Cosmology is highly useful 
\cite{K92}, and developing RPM's adds to this.

1-$d$ RPM's \cite{AF, ScaleQM} and triangleland in 2-$d$ \cite{08I, 08II, +tri, 08III} have already 
been covered.  
2-$d$ suffices for almost all the analogies with GR to hold whilst keeping the mathematics manageable: 
for 1- and 2-$d$ pure-shape RPM's the reduced configuration spaces (shape spaces) are 
$\mathbb{S}^{N - 2}$ and $\mathbb{CP}^{N - 2}$; for scaled RPM's, they are the cones over these.    
Thus, we are now looking at the case of quadrilateralland in 2-$d$.  
Quadrilateralland is valuable A) through its simultaneously possessing nontrivial subsystems and 
nontrivial constraints, rendering it a useful and nontrivial model from which to gain qualitative 
understanding of the Halliwell--Hawking model and of correlations within a single instant, which enter 
into many timeless approaches to the Problem of Time (e.g. \cite{H03, Records} and more references in 
\cite{APOT}). 
B) From the 2-$d$ RPM of $N$ particles having $\mathbb{CP}^{N - 2}$ mathematics, of which $N$ = 4 is 
the first genuinely nontrivial and thus typical case (since, for $N = 3$, $\mathbb{CP}^1 = \mathbb{S}^2$).  
I further lay out mathematics of the quadrilateral in 2-$d$ in Sec 2, and discuss some of the 
consequences and applications in the Conclusion.

Shape quantities are useful \cite{08II, AF, +tri, Cones, ScaleQM, 08III} in interpreting the  
classical and quantum solutions of RPM's and for the kinematical quantization \cite{Isham84} of RPM's 
\cite{AF, 08III}.  
For the relational triangle, a useful set of such are Dragt-type coordinates \cite{Dragt}; these are 
nontrivially realized Cartesian coordinates associated with a $\mathbb{R}^3$ surrounding the 
$\mathbb{S}^2$ shape space.    
These can furthermore be interpreted as an ellipticity of the moments of inertia, a departure 
from isoscelesness (`anisoscelesness') and four times the area per unit moment of inertia of the 
(mass-weighted) triangle (\cite{+tri} and Sec 4).
Ellipticity and anisoscelesness depend on how one splits the subsystem up into clusters, while the area 
is a clustering-independent property, alias {\it democracy invariant} \cite{ZickACG86, LR95, LR} (Sec 3).
The current paper addresses the question of what the relational quadrilateral's counterparts of 
these Dragt-type quantities are (Sec 5).   
I find that they are 3 anisoscelesnesses, 1 ellipticity, 1 linear combination of 2 ellipticities and a 
quantity proportional to the square root of the sum of squares of mass-weighted areas. 
(All of these quantities refer to Sec 2's coarse-graining triangles present within the quadrilateral.)  
Among these, the quantity related to the areas is the sole democracy invariant shape quantity.  
The current paper has useful applications to 1) investigations of the classical and quantum mechanics of the 
relational quadrilateral (paralleling those in \cite{08I, 08II, +tri, 08III} for the relational 
triangle). 
2) These are useful for subsequent investigations of Problem of Time in Quantum Gravity strategies \cite{APOT, SemiclIII, Forth} 
and for investigation of various other conceptual issues in Quantum Cosmology \cite{H03, AF, +tri, QSub, 
Forth}, including in particular  the role of democratic invariants as diagnostics of uniform states.
Uniform states play an important role in both Classical and Quantum Cosmology, since the current universe 
is fairly uniform and initial conditions from which it arose are usually taken to be extremly uniform.

\section{The relational quadrilateral}  

$N$ particles in 2-$d$ has\footnote{I refer to 
this as $N$-a-gonland, the first two nontrivial subcases of which are 
triangleland ($N$ = 3) and quadrilateralland ($N$ = 4).} 
Cartesian configuration space $\mathbb{R}^{2N}$.
Rendering absolute position irrelevant (e.g. by passing from particle position coordinates\footnote{I 
take $a$, $b$, $c$ as particle label indices running from 1 to $N$ for particle positions,
$e$, $f$, $g$ as particle label indices running from 1 to $n = N - 1$ for relative position variables,  
$I$, $J$, $K$ as indices running from 1 to $n - 1$, and
$i$, $j$, $k$ as spatial indices.
$\Delta$ is an index running over the $N\{N - 1\}/2$ entries of the symmetric $n \times n$ matrix 
that coarse-grains the $Q$-matrix in accordance with the block structure of the democracy transformations.    
(This runs from 1 to 3 for triangleland and 1 to 6 for quadrilateralland.)
$\delta$ is an index running over all bar the last of the $\Delta$ indices.
$p$, $q$, $r$ as relational space indices (running from 1 to 3 for triangleland and from 1 to 5 for 
quadrilateralland), and
$u$, $v$, $w$ as shape space indices (running from 1 to 2 for triangleland and from 1 to 4 for quadrilateralland).}
${\bq}^a$ to any sort of relative coordinates) leaves one on the configuration space {\it relative 
space}, $\fR = \mathbb{R}^{2n}$, for $n$ = $N$ -- 1.
The most convenient sort of coordinates for this are {\it relative Jacobi coordinates} \cite{Marchal} 
${\bR}^e$.  
These are combinations of relative position vectors ${\br}^{ab} = {\bq}^b - {\bq}^a$ between particles 
into inter-particle cluster vectors such that the kinetic term is diagonal.  
E.g. for 3 particles, these are ${\bR}_1 = {\bq}_3 - {\bq}_2$ and ${\bR}_2 = {\bq}_1 - \{m_2{\bq}_2 + 
m_3{\bq}_3\}/\{m_2 + m_3\}$.  
Relative Jacobi coordinates have associated particle cluster masses $\mu_e$. 
In fact, it is tidier to use {\it mass-weighted} relative Jacobi coordinates $\brho^e = \sqrt{\mu_e}
{\bR}^e$ (Fig \ref{Fig1}) or the squares of their magnitudes: partial moments of inertia $I^e = \mu_e
|{\bR}^e|^2$, or the normalized counterparts of the former, ${\bn}^e = \brho^e/\rho$. 
Here, $\rho = \sqrt{I}$ is the {\it hyperradius}, $I$ itself being the moment of inertia of the 
system.
I use (Hb) as shorthand for \{ab, cd\} i.e. the clustering (partition into subclusters) into two pairs 
ab and cd, and (Ka) as shorthand for \{\{cd, b\}, a\} i.e. the clustering into a single particle a 
and a triple \{cd, b\} which is itself partitioned into a pair cd and a single particle b.
In each case, a, b, c, d form a cycle.
I take clockwise and anticlockwise labelled triangles to be distinct, i.e. I make the plain rather than 
mirror-image-identified choice of set of shapes.
I also assume equal masses for simplicity.  

{            \begin{figure}[ht]
\centering
\includegraphics[width=1.0\textwidth]{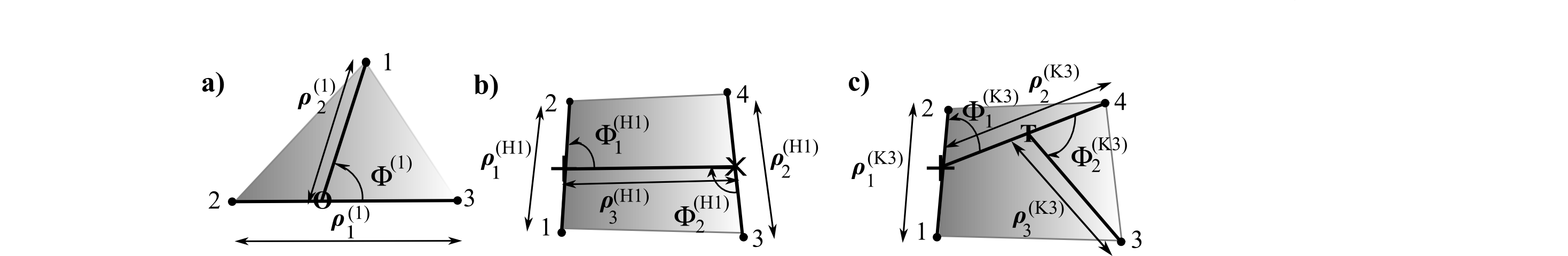}
\caption[Text der im Bilderverzeichnis auftaucht]{        \footnotesize{O, +, $\times$ and T denote COM(23), 
COM(12), COM(34) and COM(124) respectively, where COM(ab) is the centre of mass of particles a and b. 

\noindent
a) For 3 particles in the plane, one particular choice of mass-weighted relative Jacobi coordinates are 
as indicated. 
I furthermore define $\Phi^{(\sa)}$ as the `Swiss army knife' angle 
$\mbox{arccos}\big( \brho_1^{(\sa)} \cdot \brho_3^{(\sa)} / \rho_1^{(\sa)} \rho_3^{(\sa)} \big)$.

\noindent b) For 4 particles in the plane, one particular choice of mass-weighted relative Jacobi 
H-coordinates are as indicated.
I furthermore define $\Phi_1^{(\sH\sb)}$ and $\Phi_2^{(\sH\sb)}$ as the `Swiss army knife' angles 
$\mbox{arccos}\big(\brho_1^{(\sH\sb)}\cdot\brho_3^{(\sH\sb)}/\rho_1^{(\sH\sb)}\rho_3^{(\sH\sb)}\big)$ 
and  
$\mbox{arccos}\big(\brho_2^{(\sH\sb)}\cdot\brho_3^{(\sH\sb)}/\rho_2^{(\sH\sb)}\rho_3^{(\sH\sb)}\big)$ 
respectively. 

\noindent c) For 4 particles in the plane, one particular choice of mass-weighted relative Jacobi 
K-coordinates are as indicated.
I furthermore define $\Phi_1^{(\sK\sa)}$ and $\Phi_2^{(\sK\sa)}$ as the `Swiss army knife' angles  
$\mbox{arccos}\big( \brho_1^{(\sK\sa)}\cdot\brho_2^{(\sK\sa)}/\rho_1^{(\sK\sa)}\rho_2^{(\sK\sa)} \big)$ 
and  
$\mbox{arccos}\big( \brho_2^{(\sK\sa)}\cdot\brho_3^{(\sK\sa)}/\rho_2^{(\sK\sa)}\rho_3^{(\sK\sa)} \big)$ 
respectively.  
Later references to H and K coordinates refer explicitly to the (H1) and (K3) cases depicted above; I 
drop these labels to simplify the notation. }         }
\label{Fig1} \end{figure}          }

If one takes out the rotations also, one is on {\it relational space}.  
If one furthermore takes out the scalings, one is on {\it shape space}. 
If one takes out the scalings but not the rotations, one is on {\it preshape space} \cite{Kendall}.  
For $N$ particles in 2-$d$, preshape space is $\mathbb{S}^{nd - 1}$ and shape space is 
$\mathbb{CP}^{\sN - 2}$ (provided that the plain choice of set of shapes is made).  
The 3-particle case of this is, moreover, special, by $\mathbb{CP}^1 = \mathbb{S}^2$.
Now, relational space is {\it the cone} \cite{Cones} over shape space, which is 
C$(\mathbb{CP}^{\sN - 2})$ in the present case [and $\mC(\mathbb{CP}^1) = \mC(\mathbb{S}^2) = 
\mathbb{R}^3$, albeit not with the flat metric upon it \cite{08III}].  
Thus, we consider this paper's quadrilateralland case as the first case with nontrivial 
complex-projective mathematics.

In this scheme, the kinetic metric on shape space is the natural Fubini-Study metric \cite{Kendall, FORD}, 
\beq
\d s_{\sF\sS}^2 = \big\{\{1 + |{\mZ}|_{\sc}^2\}|\d {\mZ}|_{\sc}^2 - |({\mZ} ,\d\overline{{\mZ}})_{\sc}|^2\big\}/
\{1 + |{\mZ}|_{\sc}^2\}^2  \mbox{ } .
\label{FS} 
\eeq
Here, $|{\mZ}|_{\sc}^2 = \sum_{I}|{\mZ}^{I}|^2$, $( \mbox{ } , \mbox{ } )_{\sc}$ is the corresponding inner 
product, ${\mZ}^I$ are the inhomogeneous coordinates [independent set of ratios of $\mz^{e} = \rho^{e}
\mbox{exp}({i\phi^{e}})$ for ${\rho}^{e}$, $\phi^{e}$ the polar presentation of Jacobi 
coordinates], the overline denotes complex conjugate and $|\mbox{ }|$ is the complex modulus.
Note that using $\mZ^{I} = {\cal R}^{I}\mbox{exp}({i\Phi^{I}})$, the corresponding kinetic term can be 
recast in real form. 
The kinetic metric on relational space in scale-shape split coordinates is then  
\beq
\d s_{\sC(\sF\sS)}^2 = \d \rho^2 + \rho^2\d s_{\sF\sS}^2 \mbox{ } . 
\eeq
The action for $N$-a-gonland in relational form is then 

\noindent
\beq
\fS = 2\int\d\lambda \sqrt{\fT  \{\fU + \fE\} } \mbox{ } . 
\eeq
This is a Jacobi-type action \cite{Lanczos} (thus complying with temporal relationalism), with 
$\fT = \fT_{\sF\sS}$ or $\fT_{\sC(\sF\sS)}$ built from $\d s^2_{\sF\sS}$ or $\d s^2_{\sC(\sF\sS)}$ 
respectively (thus directly implementing configurational relationalism).  
$\fU$ is here minus the potential energy, and $\fE$ is the total energy.

\section{Democracy transformations for 1- and 2-d RPM's}  

The {\it democracy tranformations} \cite{ZickACG86,LR95,LR} (alias kinematical rotations \cite{ML99})
\beq
\brho_{i} \longrightarrow \brho_{i}^{\prime} = 
D_{ij}\brho_{j}
\eeq
are interpolations between relative Jacobi vectors for different choices of clustering.
These transformations form the {\it democracy group} Demo($n$), which is $O(n)$, but can be taken 
without much loss of generality to be $SO(n)$. 
Note that this is a symmetry group of the unreduced kinetic term.  
Also note that the democracy group is indeed independent of spatial dimension, since democracy 
transformations only mix up whole Jacobi vectors rather than separately mixing up each's individual  
components.  
This means, firstly, that much about the theory of democracy transformations in 2-$d$ can be gained from 
their more widely studied counterpart in 3-$d$.
Secondly, it means that for spatial dimension $> 1$, the democracy group is but a subgroup of the full 
symmetry group of the unreduced kinetic term, which itself is the dimension-dependent $O(nd)$.

\mbox{ }  

\noindent {\bf Lemma 1}. The hyperradius $\rho$ is always a democracy invariant.  
It is, however, a scale rather than shape quantity.

\mbox{ } 

\noindent {\bf Lemma 2}. 
$\mbox{dim}({\cal R}(n, d)) - \mbox{dim}(\mbox{Demo}(n)) = \{n + d - \{n - d\}^2\}/2$ gives a 
a minimum number of independent democracy invariants \cite{LR95}.
This bound is 2 both for triangleland and for quadrilateralland, revealing in each case the presence of 
at least one further democracy invariant.  

\mbox{ } 

\noindent
It is intuitively clear that one such for triangleland should be the area of the triangle. 
It is not, however, clear what happens for quadrilateralland, due to RPM's in truth involving  
constellations, i.e. sets of points in 2-$d$ rather than figures formed from these points. 
Then, for $N > 3$, there is an ambiguity in how one would `join up the dots'. 
Thus the area ascribed to the constellation via drawing a quadrilateral between 
its points is not itself a democracy invariant.

To  find democracy invariants, consider the independent invariants that 
arise from the $Q$-matrix \cite{LR95}
\beq 
Q_{ij} = \brho_i\brho_j^{\sT} \mbox{ } .  
\eeq  
The invariant trace($Q$) always gives the hyperradius, hence proving Lemma 1.  
For $N$-stop metroland, $Q_{ij} = \rho_i\rho_j$, which is sufficiently degenerate that it and all 
of its nontrivial submatrices have zero determinants, so that there are no more democracy invariants.  
For triangleland, however, det($Q$) = 4 Area$^2$ (for Area the area of the mass-weighted triangle 
per unit $I$), while $Q_{ij}$ is but 2 $\times$ 2, so that there are no more invariants.  
For quadrilateralland, det($Q$) = 0.  
This is understandable in terms of volume forms being zero for planar figures.  
Also, being 3 $\times$ 3, there is one further invariant, $Q_{11}Q_{22} - Q_{12}\mbox{}^2$ + cycles = 
$\rho_1\mbox{}^2\rho_2\mbox{}^2 - \{\brho_1\cdot\brho_2\}^2$ + cycles = $|\brho_1\cr\brho_2|_3\mbox{}^2$ + cycles, 
i.e. proportional to the sum of squares of areas of various coarse-graining triangles (see Fig 2).
Here the 3-suffix denotes that it is mathematically the component in a ficticious third dimension 
of the cross product.  
In the \{12, 34\} H-clustering, these are the coarse-grainings by taking 1, 2 and COM(34); 3, 4 and 
COM(12); 12 and 34 with one shifted so that its COM coincides with the other's.  
See \cite{QSub} for the K-clustering counterpart.  
These triangles clearly correspond to those that one makes by taking two out of the three relative Jacobi 
variables; as such, they play further roles below. 
This is also indicatory of how nontrivial subsystems have a role in quadrilateralland, which is of 
interest especially as regards timeless approaches (see the Conclusion for more).

{            \begin{figure}[ht]
\centering
\includegraphics[width=0.8\textwidth]{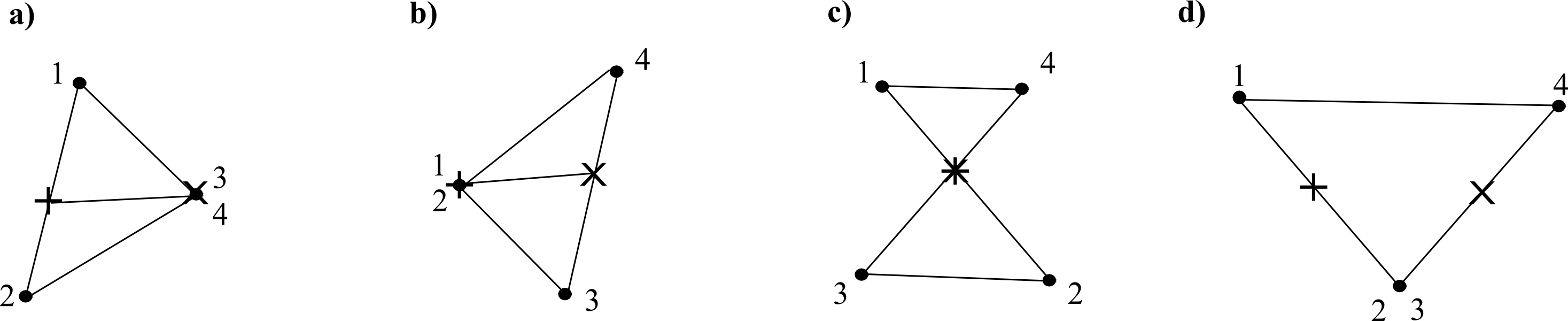}
\caption[Text der im Bilderverzeichnis auftaucht]{        \footnotesize{Figure of the coarse-graining 
triangles.  
For the (H2) coordinates for a given quadrilateral [a)], then collapsing each of 
$\rho_1^{(\tH 2)}$, $\rho_2^{(\tH 2)}$ and $\rho_3^{(\tH 2)}$ in turn gives the coarse-graining 
triangles b), c) and d).}         }
\label{Constit-Tri}\end{figure}          }

\section{Shape quantities for triangleland}

\noindent I present a different method of finding the useful Dragt-type shape coordinates for triangleland, 
a method which I find to extend to the quadrilateralland case, as well as providing a useful prequel for 
this as regards notation and interpretation.   
$\underline{\underline{Q}}$ may be written as $\frac{1}{2}\{w \underline{\underline{1}} + 
w_1\underline{\underline{\sigma}}_1 + w_3\underline{\underline{\sigma}}_3\}$ (for Pauli matrices 
$\underline{\underline{\sigma}}_1$ and $\underline{\underline{\sigma}}_3$; 
$\underline{\underline{\sigma}}_2$ does not feature since $\underline{\underline{Q}}$ is symmetric). 
This takes such an $SU(2)$ form due to the 3-particle case being exceptional through the various tensor 
fields of interest possessing higher symmetry in this case \cite{LR95}.
Then, reading off from the definition of $\underline{\underline{Q}}$ [and dropping (1)-labels], 

\noindent
\beq
w = \rho^2 = I ( = \mbox{moment of inertia) := Size} \mbox{ } , \mbox{ } 
\eeq
\beq
w_1 = \rho_1^2 - \rho_2^2 := \mbox{Ellip} \mbox{ } , \mbox{ }  
\eeq
\beq
w_3 = 2\,\brho_1\cdot\brho_2 := \mbox{Aniso} \mbox{ } . 
\eeq
$w_1$ can be interpreted as an `ellipticity', being the difference in the principal moments of inertia 
of the base and median.  
Also, $w_3$ can be interpreted as an `anisoscelesness' (i.e. departure from isoscelesness, in analogy
with anisotropy as a departure from isotropy in GR Cosmology); specifically, working in mass-weighted 
space, Aniso per unit base length is the amount by which the perpendicular to the base fails to bisect 
it ($l_1 - l_2$ from Fig 3).

{            \begin{figure}[ht]
\centering
\includegraphics[width=0.2\textwidth]{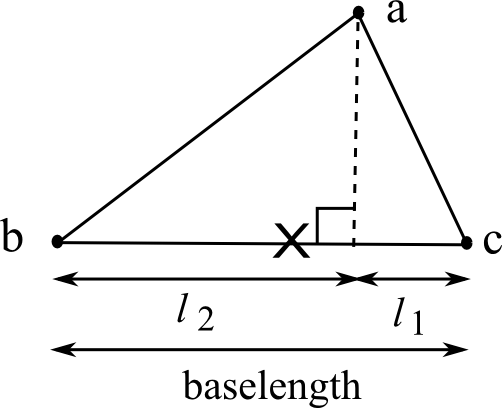}
\caption[Text der im Bilderverzeichnis auftaucht]{        \footnotesize{ Definitions of $l_1$ and $l_2$ 
used in the anisoscelesness notion Aniso(a).                         }        }
\label{Fig-3tri}\end{figure}            }

There is also a 
\beq
w_2 = 2\,|\brho_1 \cr \brho_2|_3  
\label{9}
\eeq 
such that $w^2 = \sum\mbox{}_{\mbox{}_{\mbox{\scriptsize i = 1}}}^3w_i\mbox{}^2$.  
(\ref{9}) has straightforward interpretation as 4 $\times \mbox{ Area}$, which is a measure of 
noncollinearity and, clearly, a democratic invariant notion.

For triangleland, $w_1$, $w_2$ and $w_3$ are Dragt-type coordinates. 
[Though, given that these run over not Dragt's $\mathbb{R}_+^3$ but $\mathbb{R}^3$, they are even more 
closely related to the well-known $\mathbb{S}^3 \longrightarrow  \mathbb{S}^2 $ Hopf Map.  
This is involved in the configuration space study as follows: relative space $\mathbb{R}^4 
\longrightarrow$ preshape space $\mathbb{S}^3 \stackrel{\mbox{\scriptsize Hopf map}}{\mbox{\Large $\longrightarrow$}}$ 
shape space $\mathbb{S}^2 = \mathbb{CP}^1 \stackrel{\mbox{\scriptsize coning}}{\longrightarrow} 
\mathbb{R}^3$ relational space.]
For these to be shape quantities, I divide each of these by $I$ and switch the 2 and 3 labels (so the 
principal axis is aligned with the democracy invariant), obtaining  
\beq
s_1 = \mn_1^2 - \mn_2^2    = \mbox{Ellip}/I         := \mbox{ellip} 
\mbox{ } , \mbox{ } 
\eeq
\beq
s_2 = 2\,\bn_1\cdot\bn_2   = \mbox{Aniso}/I         := \mbox{aniso} 
\mbox{ } , \mbox{ } 
\eeq
\beq
s_3 = 2|\bn_1 \cr \bn_2|_3 = 4 \times \mbox{Area}/I := 4 \times \mbox{area} := 
\mbox{demo}(N = 3) 
\mbox{ } . \mbox{ }
\eeq
These are in $\mathbb{R}^3 = C(\mathbb{S}^2)$ and subject to the on-$\mathbb{S}^2$ restriction 
$\sum\mbox{}_{\mbox{}_{\mbox{\scriptsize $\Delta$ = 1}}}^3s_{\Delta}\mbox{}^2 = 1$.
Interpretation of various great circles and of the hemispheres they divide triangleland into are 
provided in Fig 4.

{            \begin{figure}[ht]
\centering
\includegraphics[width=0.5\textwidth]{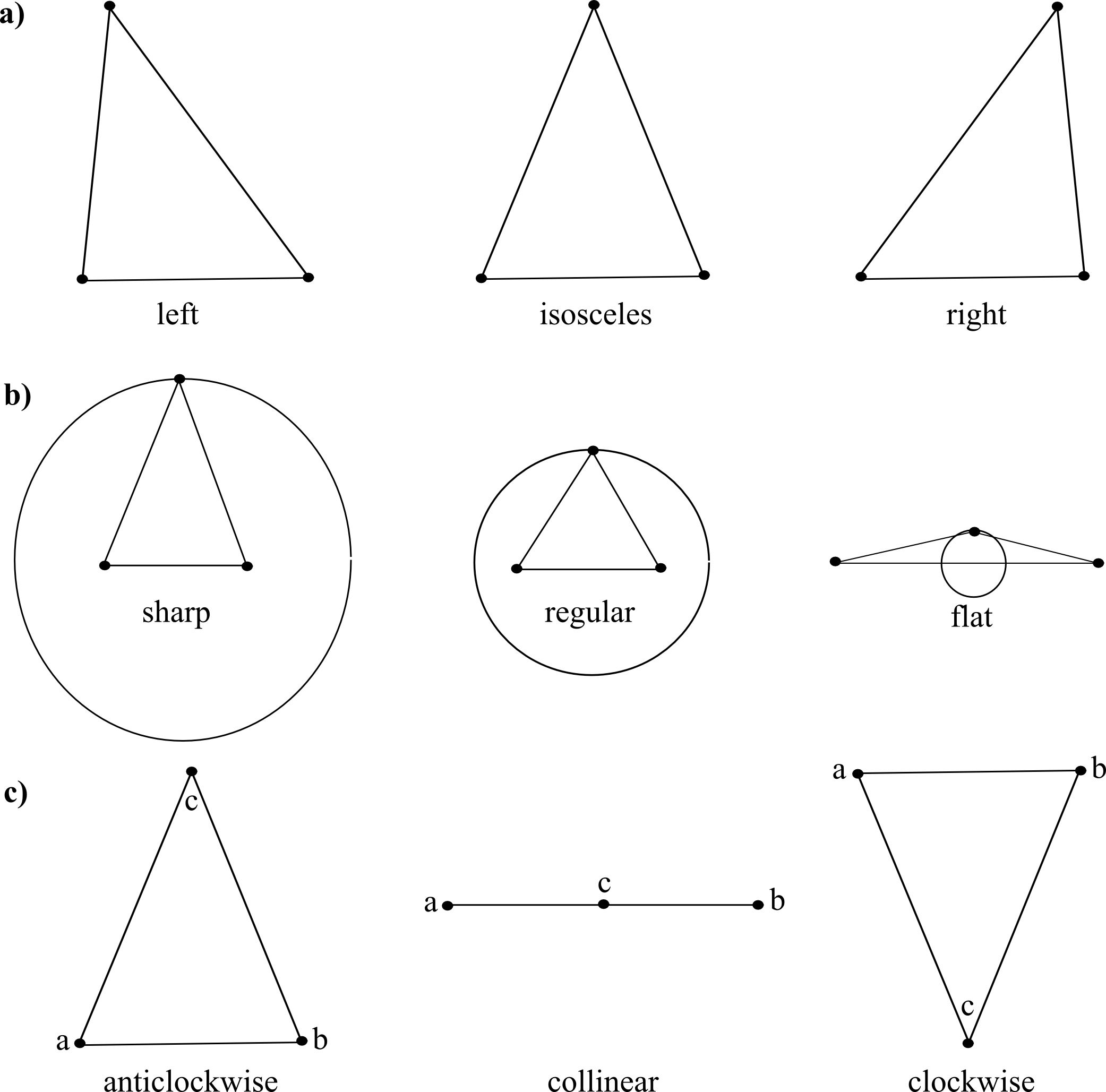}
\caption[Text der im Bilderverzeichnis auftaucht]{        \footnotesize{ 
\noindent a) Each bimeridian of isoscelesness cuts the shape space sphere into hemispheres of 
right and left triangles.  

\noindent b) Each bimeridian of regularness ($I_1 = I_2$) cuts the shape space sphere into hemispheres 
of sharp ($I_1 < I_2$)  and flat triangles ($I_1 > I_2$).  

\noindent c) The equator of collinearity cuts the shape space sphere into hemispheres of 
anticlockwise and clockwise triangles.                        }        }
\label{LRTFAC}\end{figure}            }

\vspace{10in}

\section{Shape quantities for quadrilateralland}

Repeating this for quadrilateralland closely parallels the analysis for the tetrahaedron in \cite{LR95} 
because the democracy group does not care about dimension. 
Now, 
\beq
\underline{\underline{Q}} = \frac{1}{2}
\left\{
w \,\underline{\underline{1}} + \sum\mbox{}_{\mbox{}_{\mbox{\scriptsize $\delta$ = 1}}}^5 w_{\delta}\, 
\underline{\underline{B}}_{\delta}\right\} \mbox{ } , 
\eeq
where the $\underline{\underline{B}}_{\Delta}$ are $\mj = 1$  representation matrices of the $SO(3)$ 
democracy group (they are also proportional to the real symmetric subset of the Gell--Mann $\lambda$-matrices),  
\beq
\stackrel{\mbox{$\underline{\underline{B}}_1 = \sqrt{3}$}}{} \mbox{\Huge (}
\stackrel{                \mbox{\scriptsize{1 \, 0 \, 0}} \hspace{0.03in}   }
         {    \stackrel{  \mbox{\scriptsize 0   -1 \, 0   }    }
                       {  \mbox{\scriptsize 0 \, 0 \, 0      }    }    } \mbox{\Huge )} 
\mbox{ } , \mbox{ } 
\stackrel{\mbox{$\underline{\underline{B}}_2 = \sqrt{3}$}}{}   \mbox{\Huge (}
\stackrel{                \mbox{\scriptsize{0 \,  1 \, 0}}    }
         {    \stackrel{  \mbox{\scriptsize 1 \, 0 \,  0   }    }
                       {  \mbox{\scriptsize 0 \, 0 \,  0  }    }    } \mbox{\Huge )}\mbox{ } , \mbox{ }
\stackrel{\mbox{$\underline{\underline{B}}_3 = \sqrt{3}$}}{}   \mbox{\Huge (}
\stackrel{    \mbox{\scriptsize{0 \, 0 \, 0}}    }
         {    \stackrel{  \mbox{\scriptsize 0 \, 0 \, 1   }    }
                       {  \mbox{\scriptsize 0 \, 1 \, 0  }    }    } \mbox{\Huge )}\mbox{ } , \mbox{ }
\stackrel{\mbox{$\underline{\underline{B}}_4 = \sqrt{3}$}}{}   \mbox{\Huge (}
\stackrel{    \mbox{\scriptsize{0 \, 0 \, 1}}    }
         {    \stackrel{  \mbox{\scriptsize 0 \, 0 \, 0   }    }
                       {  \mbox{\scriptsize 1 \, 0 \, 0   }    }    } \mbox{\Huge )}\mbox{ } , \mbox{ }
\stackrel{\mbox{$\underline{\underline{B}}_5 =$}}{}     \mbox{\Huge (}
\stackrel{    \mbox{\scriptsize{-1 \, 0 \, 0}}    }
         {    \stackrel{  \mbox{\scriptsize \, 0 -1  \, 0 }    }
                       {  \mbox{\scriptsize \, 0 \, 0 \,  2  }    }    } \mbox{\Huge )}\mbox{ } . \mbox{ }
\eeq
Thus, reading off from the definition of $\underline{\underline{Q}}$ and dropping (H2) labels, 
\beq
w = Q_{11} + Q_{22} + Q_{33} = \rho_1\mbox{}^2 + \rho_2\mbox{}^2 + \rho_3\mbox{}^2 = \rho^2 = I \mbox{ : moment of inertia (MOI)} \mbox{ } , 
\eeq
\beq
w_1 = \sqrt{3}\,\{Q_{11} - Q_{22}\}/2 = \sqrt{3}\{\rho_1\mbox{}^2 - \rho_2\mbox{}^2\}/2 = \sqrt{3}\,\mbox{Ellip}(12)/{2} 
\mbox{ } ,
\eeq
\beq
w_2 = \sqrt{3}\,Q_{12} =  \sqrt{3}\,\brho_1\cdot\brho_2 = \sqrt{3}\,\mbox{Aniso}(12) \mbox{ } , 
\eeq
\beq
w_3 = \sqrt{3}\,Q_{23} =  \sqrt{3}\,\brho_2\cdot\brho_3 = \sqrt{3}\,\mbox{Aniso}(23) \mbox{ } , 
\eeq
\beq
w_4 = \sqrt{3}\,Q_{31} =  \sqrt{3}\,\brho_3\cdot\brho_1 = \sqrt{3}\,\mbox{Aniso}(31) \mbox{ } , 
\eeq
\beq
w_5 = \{- Q_{11} - Q_{22} + 2Q_{33}\}/2 = \{- \rho_1\mbox{}^2 - \rho_2\mbox{}^2 + 2\,\rho_3\mbox{}^2\}/2 
    = \{\mbox{Ellip}(31) + \mbox{Ellip}(32)\}/2  \mbox{ } .
\eeq
Here, the brackets refer to the pair of Jacobi vectors involved in each coarse-graining triangle (Fig 2); 
note that Ellip being a signed quantity requires this to be an ordered pair.

Note also that if one sets 
\beq
w_6 = \sqrt{3}\{|\brho_1\cr\brho_2|_3\mbox{}^2 + \mbox{cycles}\}^{1/2} 
= \sqrt{12}\{\mbox{Area}(12)^2 + \mbox{cycles}\}^{1/2},
\eeq 
i.e. a quantity proportional to the square root of the above-found democracy invariant, then 
$w^2 = \sum\mbox{}_{\mbox{}_{\mbox{\scriptsize $\Delta$ = 1}}}^6w_{\Delta}^2$.
One can then straightforwardly make the $w_{\Delta}$ into surrounding shape quantities by dividing through by $I$: 
\beq
s_1 := \sqrt{3}\{\mn_1\mbox{}^2 - \mn_2\mbox{}^2\}/2 = \sqrt{3}\,\mbox{ellip}(12)/{2}
\eeq
\beq
s_2 := \sqrt{3}\,{\bn_1}\cdot{\bn_2} = \sqrt{3}\,\mbox{aniso}(12) \mbox{ } ,
\eeq
\beq
s_3 := \sqrt{3}\,{\bn_2}\cdot{\bn_3} = \sqrt{3}\,\mbox{aniso}(23) \mbox{ } ,
\eeq
\beq
s_4 := \sqrt{3}\,{\bn_3}\cdot{\bn_1} = \sqrt{3}\,\mbox{aniso}(31) \mbox{ } ,
\eeq
\beq
s_5 := \{- \mn_1\mbox{}^2 - \mn_2\mbox{}^2 + 2\,\mn_3\mbox{}^2\}/2 = \{\mbox{ellip}(31) + \mbox{ellip}(32)\}/2 \mbox{ } ,
\eeq
\beq
s_6 := \sqrt{3}\{|\bn_1 \cr \bn_2|_3\mbox{}^2 + \mbox{cycles}\}^{1/2} 
= \sqrt{12}\{\mbox{area}(12)^2 + \mbox{cycles}\}^{1/2} = \mbox{demo}(N = 4) \mbox{ } .  
\eeq
These belong to $\mathbb{R}^6 = \mathbb{C}^3$ and the relation between the $w^{\Delta}$ becomes the 
on-$\mathbb{S}^5$ condition $\sum\mbox{}_{\mbox{}_{\mbox{\scriptsize $\Delta$ = 1}}}^6
s_{\Delta}\mbox{}^2 = 1$. 
That an $\mathbb{S}^5$ makes an appearance is not unexpected (for quadrilateralland, from Sec 2  
and using the generalization af the Hopf map in the sense of $U(1) = SO(2)$ fibration, 
relative space $\mathbb{R}^6 \longrightarrow$ preshape space $\mathbb{S}^5 
\stackrel{\mbox{\scriptsize Hopf map}}{\mbox{\Large $\longrightarrow$}}$ shape space $\mathbb{CP}^2$).   
Moreover, since $\mathbb{CP}^2$ is 4-$d$ and $\mathbb{S}^5$ is 5-$d$, there is a further condition on 
the $s_{\Delta}$ for quadrilateralland.  
Also above, aniso(12) is same notion of anisoscelesness as before but now applied to the triangle made 
out of the 1 and 2 relative Jacobi vectors, and so on.   
Finally, note the interpretation 
\beq
\mbox{demo($N$ = 4) = $\sqrt{\mbox{12 $times$ sum of squares of mass-weighted areas of coarse-graining triangles per unit 
MOI}}$} \mbox{ } .   
\eeq

Quadrilateralland's relational space is locally $\mathbb{R}^5$.  
One can envisage this from $\mathbb{CP}^2/\mathbb{Z}_2$ = $\mathbb{S}^4$ \cite{Kuiper} and then taking the cone.  
As we are treating $\mathbb{CP}^2$, we have, rather, two copies of $\mathbb{S}^4$.  
Such a doubling does not appear in the coplanar boundary of \cite{LR95}'s study of 4 particles in 3-$d$, 
since 3-$d$ dictates that the space of shapes on its coplanar boundary is the mirror-image-identified 
one.
Another distinctive feature of 3-$d$ is that it has a third volume-type democracy invariant, 
$\brho_1 \cdot \brho_2 \cr \brho_3$.

Finally, demo($N$ = 4) = $s_6$ can be used to diagnose collinearity as the points at which this is 
zero.  
It is in this respect like triangleland's demo($N$ = 3) = dra$_2$, which also gives the equilateral 
configurations (corresponding to a manifest notion of maximum uniformity) as the output to its 
extremization, corresponding to the natural $\mathbb{S}^2$ poles.  
However, one can quickly see that extremizing demo$(N = 4)$ will involve something more complicated, 
since there are not two but six permutations of labelled squares (likewise a manifest notion of maximum 
uniformity), so these cannot possibly all lie on the two $\mathbb{S}^5$ poles, nor could any of them 
lie on the poles by a symmetry argument.    
In fact, extremizing demo($N$ = 4) does yield these squares, but not uniquely.  
One gets, rather, one curve of extrema for each of the two orientations, each containing three labelled squares.

\section{Conclusion}

In this paper, I showed how to extend the relational triangle's useful Dragt-type shape quantities 
(ellipticity, anisoscelesness and four times the mass-weighted area per unit $I$) to the case of the relational quadrilateral.
This gives rise to one ellipticity, one linear combination of ellipticities, three anisoscelesnesses and a quantity 
proportional to the square root of the sum of the squares of the three mass-weighted areas per unit $I$.
These quantities refer to three cluster-dependent `coarse-graining triangles' (Fig 2). 
The last of these quantities, is furthermore like the triangle case's 4 $\times$ area [= demo($N$ = 3)] in being a  
democratic invariant (cluster independent quantity), demo($N$ = 4).

In 2-$d$, the extension to the general `$N$-a-gonland' is straightforward to start off with because these have $\mathbb{CP}^{N - 2}$ 
shape spaces; this is in contrast to the 3-$d$ models' shape spaces forming a much harder 
sequence \cite{Kendall}, which is a good reason to study the 2-$d$ models (particularly since the 
analogy with geometrodynamics does not require dimension 3 in order to work well).  
There are now $n$ - 1 independent ellipticities, one hyperradius $\rho$, one  
sum of squares of areas type of democracy invariant, but, additionally,   
the number of anisoscelesnesses goes as $n$\{$n$ -- 1\}/2, so that the latter substantially overwhelms 
the shape space dimension.
This suggests that the these more general models will require a tighter set of shape variables than 
the present paper's one which includes all of its possible anisoscelesness quatities.  
I leave this (and its application to kinematical quantization) as questions for a further occasion.

This paper's shape quantities are useful in the following investigations of the classical and quantum 
mechanics of the relational quadrilateral \cite{QSub, Forth} (paralleling those in \cite{08I, 08II, +tri, 
08III} for the relational triangle, including expectations and spreads of corresponding operators).

\noindent 
A) I look to study the special configurations and subregions of quadrilateralland's shape space that 
they represent \cite{QSub}.  
I find there that a simple linear recombination of the present paper's \{$I$, $s_{\delta}$\} coordinates 
are useful for this purpose.  

\noindent
B) I physically interpret quadrilateralland's conserved quantities (these form the group $SU(3)$ as 
opposed to the $SU(2)$ and $SO(3)$ forms of triangleland \cite{+tri} and 4 particles on a line 
\cite{AF} respectively).  
I then aim to treat quadrilateralland's classical equations of motion and Schr\"{o}dinger equation 
in forthcoming papers.

\noindent 
C) These are useful for subsequent investigations of Problem of Time in Quantum Gravity strategies 
and various other quantum-cosmological issues.  
Particular such application for Quadrilateralland and $N$-a-gonland are to 
timeless approaches to the Problem of Time in Quantum Gravity, 
qualitative models of structure formation in Quantum Cosmology, and 
to the robustness study based on the \{$N$ -- 1\}-a-gon model lying inside the 
$N$-a-gon one.   
I note that complex projective mathematics (the present paper involving the simplest RPM model paper 
with nontrivial such) will underly this robustness study. 

\noindent 
The large-$N$ limit of $N$-a-gonland may also be interesting, alongside the study of the statistical 
mechanics/entropy/notions of information/relative information/correlation that timeless approaches are 
concerned with.   
The present paper's model is also the smallest that possesses both linear rotational constraints 
(conferring it a `midisuperspace toy model character') and nontrivial subsystems (applications of which 
include attempting to use one such as a clock for the other parts of the model universe \cite{PW83, 
Rovellibook}, timeless records approaches \cite{Records} and structure formation.)  

\noindent Another application is to notions of uniformity and how probable such states are (this is 
another issue of interest in Quantum Cosmology).
Now, while demo($N = 4$) for quadrilateralland is as good a diagnostic as demo($N$ = 3) for 
triangleland as regards collinear configurations, that the extrema of demo($N$ = 3) pick out the 
triangleland uniform states (the 2 labellings of equilateral triangle) only partly carries over to the 
extrema of demo($N$ = 4), since these do not uniquely pick out the quadrilateral's most uniform states (the 6 
labellings of the square). 
Obtaining more information of where these uniform states sit in quadrilateralland's $\mathbb{CP}^2$ 
configuration space (with or without orientation and distinguishability reducing these to 3 or just 1 
uniform state) remains work in progress \cite{QSub}.  

\mbox{ }
 
\noindent {\bf Acknowledgements}: I thank: my wife Claire, Amelia, Sophie, Sophie, Adam, Sally and Tea for 
being supportive of me whilst this work was done.   
Professors Belen Gavela, Karel Kucha\v{r}, Marc Lachi\`{e}ze-Rey, Malcolm MacCallum, Don Page, Reza 
Tavakol, and Dr's Julian Barbour and Jeremy Butterfield for support with my career.  
Fqxi grant RFP2-08-05 for travel money, and Universidad Autonoma de Madrid for funding during part of 
this work.


\end{document}